\documentclass[12pt, a4paper]{article}
\usepackage{amssymb,amsmath, amsthm}
{

  }

\usepackage[pdftex]{color}
\usepackage{chngcntr}
\usepackage[utf8]{inputenc}
\usepackage{pdfsync}
\usepackage{authblk}
\usepackage[toc,page]{appendix}
\usepackage[
backend=biber,
style=authoryear,
citestyle=authoryear,
maxbibnames=6,
maxcitenames=2
]{biblatex}
\usepackage{titlesec}
\usepackage[a4paper,bottom=3cm,left=2.5cm,right=2.5cm]{geometry}
\usepackage{graphicx}
\usepackage{hyperref}
\usepackage{float}
\restylefloat{table}
\usepackage{multirow}
\usepackage{caption} 
\usepackage[table]{xcolor}
\graphicspath{ {C:/Users/Alessandra/Documents/Latex/immagini/} }
\numberwithin{equation}{section}
\numberwithin{lemma}{section}
\numberwithin{theorem}{section}
\numberwithin{definition}{section}
\numberwithin{remark}{section}
\numberwithin{corollary}{section}
\numberwithin{proposition}{section}
\titleformat{\paragraph}
{\normalfont\normalsize\bfseries}{\theparagraph}{1em}{}
\titlespacing*{\paragraph}
{0pt}{3.25ex plus 1ex minus .2ex}{1.5ex plus .2ex}

\begin{document}
\date{}
\title{Long-Term Unemployed hirings: Should targeted or untargeted policies be preferred?}
\author[*]{Alessandra Pasquini}
\author[**]{Marco Centra}
\author[***]{Guido Pellegrini}
\affil[*]{\small MEMOTEF,  Sapienza Universit\`{a} di Roma, \texttt{alessandra.pasquini@uniroma1.it}}
\affil[**]{\small INAPP, \texttt{m.centra@inapp.org}}
\affil[***]{\small Dipartimento di Scienze Sociali ed Economiche,  Sapienza Universit\`{a} di Roma, \texttt{guido.pellegrini@uniroma1.it}}
\pagestyle{myheadings}
\markright{\sc }
\maketitle

\abstract{
To what extent, hiring incentives targeting a specific group of vulnerable unemployed (i.e. long-term-unemployed) are more effective, with respect to generalised incentives (without a definite target), to increase hirings of the targeted group? Are generalized incentives able to influence hirings of the vulnerable group? Do targeted policies have negative side effects too important to accept them? Even though there is a huge literature on hiring subsidies, these questions remained unresolved. We tried to answer them, comparing the impact of two similar hiring policies, one oriented towards a target group and one generalised, implemented on the italian labour market.  We used administrative data on job contracts, and counterfactual analysis methods. The targeted policy had a positive and significant impact, while the generalized policy didn’t have a significant impact on the vulnerable group. Moreover, we concluded the targeted policy didn’t have any indirect negative side effect.
\bigskip
}
\vspace{2mm}


\par

\section{Introduction}\label{puffu}
In spite of the huge amount of international literature on hiring and wage subsidies, targeting a vulnerable category of unemployed (i.e. long-term-unemployed), few is known about the difference between the last and generalised subsidies without a definite target. Moreover, few is known about the mechanism behind them and their indirect effects. Additional informations on the difference between the two policies would, nontheless, be extremely useful. Indeed, it is important to know which of the two policies would be more effective and whether the vulnerable group would be penalised by a switch from the targeted to the untargeted ones. Equally, it would be useful to have more informations on their possible indirect effects, which may affect their effectiveness. Our study tries to overcome this lack of informations. We estimated the impacts of a targeted policy, and a generalised one with similar characteristics, from the targeted group perspective. Later on, we compared them. Both of the policies were implemented in Italy. The first policy is Law 407/90. According to \textcite{Bro2012} classification, it was an incentive to create (permanent) employment targeting long-term unemployed.  It was implemented from 1990 to the end of 2014. It was widely used by firms. The second policy is 2015 Legge Stabilità (or Law 190). It consisted, among others, in incentives to create permanent employment without a particular targeted group. It lasted one year. It implied the permanent end of Law 407/90. \par

There are no empirical studies comparing targeted and generalised subsidies. Nevertheless, \textcite{Bro2015} did a meta-analysis trying to identify the positive and negative consequences of both types of policies. The author underlined as targeted hiring subsidies have lower deadweigth costs\footnote{With this term the author identifies the costs of subsidies benefitting individuals who would have been hired indipendently from policy implementation.} with respect to untargeted ones. Nevertheless, he stated that targeted subsidies may have negative consequences at a macroeconomic level. Indeed, in presence of the subsidy, firms may hire subsidised unemployed instead of unsubsidised ones with similar characteristics (and, consequently, a similar vulnerability level), penalising the last. This indirect effect is called displacement effect. Moreover, when eligibility is defined by determined conditions (as in tagreted policies), the presence of asymmetric informations on Government's side may induce agents to cheat in order to appear eligibles when they're not. \textcite{Bro2015} called this indirect effect asymmetric informations effect. Our results show that the positive effects of targeted policies are strong and significant, while there are no displacement and asymmetric information effects. \\

As mentioned before, even though, to the best of our knowledge, there are no empirical evidences on the differences between targeted and untargeted policies, international literature on hiring and wage subsidies targeting vulnerable groups of individuals is huge. Literature reviews generally agree on the positive effect of this type of active labour market policies, both at macroeconomic level (\cite{Mar2015}) and at microeconomic level (\cite{Bro2015}, \cite{Cal2002}), especially after negative economic shocks (\cite{Mar2001}). One of the reasons of their success is their closeness to regular employment (\cite{Cal2002}). Most of these studies warn on the importance of a proper choice of the target group and of subsidy assignment rules (\cite{Bro2015}). The last have to be a good balance in the trade off between the achievement of a maximum net effect (which requires stricter rules) and the mantainance of a high take-up rate, which can be lowered if firms have to face too much bureaucracy (\cite{Mar2001}). Furthermore, the studies underline that a targeted policy may have multiple indirect side effects, such as displacement and asymmetric information effects (\cite{Bro2012}, \cite{Cal2002}). About single study analysis, most of the studies concluded hiring subsidies have a positive and significant effect (\cite{Sia2008}, \cite{Ann2008}, \cite{Ana2013}, \cite{Ber2008}, \cite{For2004}, \cite{Kat1998}, \cite{Mor2003}), few of them concluded they have a null effect (\cite{Sch2013}, \cite{Jae2011}, \cite{Boo2004}) and only two of them concluded they have a negative effect due to targeted group stigmatization (\cite{Klu2007}, \cite{Bur1985}). \\

The international literature study most closely related to our framework is \textcite{Sch2013}. They studied the intention-to-treatment effect of an employer-side wage subsidy program implemented in Germany. Their outcome was the exit rate from unemployment to unsubsidised employment. The program targeted long-term unemployed and it was implemented from 1989 to 2002. They exploited the 12 months of unemployment threshold defining eligibility, to apply a regression discontinuity design. Moreover, they exploited the end of the policy in 2003 to combine the regression discontinuity design with a diff-in-diff model and clean the estimation from the influence of other incentives applied at the same eligibility threshold. They found no significant effect. Even though, as the authors mentioned, the estimation may lack of some statistical power due to the low take-up rate of the policy, they underlined how this result was crucially different from \textcite{Sia2008} ones. The last studied the impact of six different swedish active labour market programmes. One of them consisted in job subsidies targeting long-term unemployed 20 to 25 years old. Using a propensity score matching, she concluded entering in the program, rather than being unemployed, increased employment rates by 35\%. \textcite{Sch2013} attributed the difference between the results to the fact that a matching method, in their opinion, wasn't enough to take into account of all the selection into treatment. Our result doesn't fit with this hypothesis. Indeed, using the same counterfactual method as \textcite{Sch2013} we detected a positive and significant effect. In contrast with \textcite{Sch2013} opinion, matching procedure has been one of the most used in the estimation of hiring subsidies impact. Among other studies using this methodology, there is \textcite{Klu2007}. The authors studied two active labour market policies implemented in Poland between 1992 and 1996. Among them, there was a job subsidized program targeting particularly disadvantaged unemployed. They matched treated and control groups with respect to demographic informations and the outcome in the twelve months preceeding the treatment. The authors found a negative and significant effect on the treated and attributed it to the stigmatization that may be associated to eligibility. Similarly, \textcite{Bur1985} attributed the negative impact detected in his study to stigmatization. Exploiting a randomized experiment implemented in Dayton, Ohio from 1980 to 1981, he evaluated the impact of vouchers provided to a random sample of disadvantaged unemployed. The employers hiring the eligibles could use the vouchers to have a portion of their wage payed. The author underlined as the end of the program after six months of implementations may have biased the results. The two policies analysed by \textcite{Klu2007} and \textcite{Bur1985}, targeted a particularly disadvantaged group of unemployed. It is unlikely for the policy we studied to have a stigmatization effect. Indeed, the target is all long-term unemployed and consequently, being eligible doesn't give any additional informations on unemployed characteristics.  \\

Among the studies estimating the impact of hiring subsidies, few of them took into account of the possible indirect effects of the policies (\cite{Sch2013}, \cite{Cal2002}, \cite{Ann2008}, \cite{Boo2012}). \textcite{Sch2013} found no significant displacement effect. Nevertheless, the policy they were studying had no significant effect either. \textcite{Cal2002} and \textcite{Ann2008}, instead, found a strong and significant displacement effect. The first work is a literature review of different empirical studies. Nevertheless, they  present a single study trying to estimate the displacement effect. It consisted in a survey to firms, hence its results should be taken with caution. The second work, instead, is based on a theoretical model, relying on strong assumptions. Finally, \textcite{Boo2012} tested for the possibility that firms post-poned some hirings in order to get the subsidy and concluded that wasn't the case. \\ 

The italian literature on hiring subsidies targeting disadvantaged categories of individuals is significantly smaller. \textcite{Ana2013} studied the impact of a subsidy incentivising the switch from a temporary to a permanent contract and targeting women and under 30 years old youths. The authors exploited the age threshold to apply a regression discontinuity design. They found a positive and significant effect. Nevertheless, their analysis was focused only on Veneto region. Notwithstanding the low number of studies on italian targeted hiring subsidies, there is a significant amount of literature which estimated the impact of the generalised ones. In particular, the incentives provided according to 2015 Legge Stabilità. Among them, \textcite{Cen2016} used a diff-in-diff model to determine the impact of the policy on the incidence of permanent contracts among the total.  The authors exploited the fact that, individuals whose previous permanent contract ended less than six months before, were not eligible for the incentive. They added some covariates to the model to make the parallelism assumption more plausible. Moreover, they estimated the outcome of the control group in 2015 using interrupted time series analysis to clean the estimation from a possible displacement effect of the policy. They found a positive and significant effect. Similar results were found by \textcite{Ses2016}. The authors estimated the impact of the same policy on different outcomes, among which the probability (both for employed and unemployed) to find a permanent job. They used a diff-in-diff model with individual, monthly and annual fixed effects. None of these studies considered 2015 Legge Stabilità from a LTU perspective. Hence, our work tries to contribute to multiple issues and to fill multiple gaps in the literature on Active Labour Market Policies (ALMPs). To the international literature, being the first study analising empirically the difference between targeted and untargeted policies and giving a reliable analysis of the indirect effects of targeted policies. To the italian literature, being the first study analising Law 407/90 and the first considering Law 190 from a LTU perspective. \\

We exploited an administrative micro-database, namely CICO database, provided by the Ministry of Labour and Social Policies. We had access to these data thanks to an agreement between Dipartimento di Scienze Sociali ed Economiche of Sapienza University of Rome and INAPP (Istituto Nazionale per l'Analisi delle Politiche Pubbliche). The database contains reliable informations on contracts' stipulated from 2008 to 2016 for a sample of individuals. Contracts informations are communicated to the Ministry directly from the employers. \\
Following \textcite{Sch2013}, we applied a regression discontinuity design to estimate the impact of Law 407/90. We used the days of unemployment as a forcing variable and added daily fixed effects to the standard model. In order to select the bandwidth, we applied the method proposed by \textcite{Cat2015}, modified in order to take into account of the daily fixed effects. We partially followed \textcite{Sch2013} to check for the presence of indirect effects. I.e., we compared the differences in hirings before and after the policy ending, for values of the forcing variable far and close to the threshold. Finally, we applied a regression discontinuity design using time as a forcing variable, to estimate the impact of the generalised incentives provided by 2015 Legge Stabilità. In order to parcel out the impact of the generalised incentives from the impact of Law 407/90 ending we used a control group to do this last estimation selected following the methodology used in the first estimation. Under the first estimation assumptions, the effect estimated on the control group, can be a good representation of the impact, on the treated, of the generalised incentives in absence of Law 407/90 ending. \\

From the analysis, it emerged Law 407/90 had a strong, positive and significant effect on LTU hirings. The estimated impact was meaningfully higher than in previous studies using eligibility as treatment. This may be due to the high take-up rate and to country characteristics. We didn't detect any displacement or asymmetric information effects. \\
The remainder of the paper is organised as follow: in section 2 are provided informations on the causes of LTU duration dependence, the role of ALMPs contrasting them, the studied policies and their comparison. In section 3 the data and the used sample are presented. In section 4 the methodologies used in each estimation are described, the needed assumptions are discussed and the results are presented. In section 5 we resumed the conclusions of this study. 

\section{Long-Term Unemployed Conditions and Italian Policies}\label{pincopallo}
\subsection{Long-Term Unemployed Stigma}
Long-term unemployment status is characterised by a strong duration dependence. Many authors (\cite{Mus2008}, \cite{Mei2016}, \cite{Hec1980}, \cite{Far2015}, \cite{Due2016}) estimated the probability to exit from unemployment status for different unemployment durations. Indipendently from the considered Country and period, they found that, once taken into account of heterogeneity, such probability lowered as the duration of unemployment increased. Past literature (\cite{Far2015}, \cite{Bro2012}, \cite{Bro2015}) identified three main causes for duration dependence:\par
\begin{enumerate}
\item The possible loss of human capital and the lack of recent working experience due to the absence from employment, whose consequence is a lower desiderability of the long term unemployed.
\item The reduction of contacts the unemployed has with the labour market. Indeed, employers often use informal channels to recruit. This reason is particularly effective during recession periods since using informal channels to recruit allows employer to reduce recruitment costs. 
\item The scarring effect affecting LTU. In real labour market there is imperfect information. Employers tend to exploit the limited tools they have, to determine whether a candidate will be productive or not. One of the tools they use is whether the candidate has been rejected by previous employers. Rejection is used as a signal of low productivity or presence of other unobservable negative characteristics. LTU are automatically classified as rejected several times (even though this is not necessary true). Many authors tried to detect and quantify this last effect. All of them concluded there was a scarring effect due to long-term unemployment (\cite{Bie2008}, \cite{Obe2008}, \cite{Omo1997}, \cite{Kro2013}, \cite{Bae2014}, \cite{Ayl2013}).
\end{enumerate}
Ideally, ALMPs lowering labour cost, as hiring subsidies, overcome these issues both in the short and long term. Indeed, the lower labour cost should counteract the lower desiderability of the LTU. Once the last  has been hired, she will recover recent working experience, increase contacts with the labour market and be screened by the hiring employer overcoming the imperfect informations characterising labour market. \\
In light of what we've just explained, we can imagine the demand curves of LTU and short-term unemployed with, on average, similar characteristics, to be representable by the following graph. The red line represents the demand curve of short-term unemployed and the blue line represents the demand curve of long-term unemployed:  \par
\includegraphics[scale=0.6]{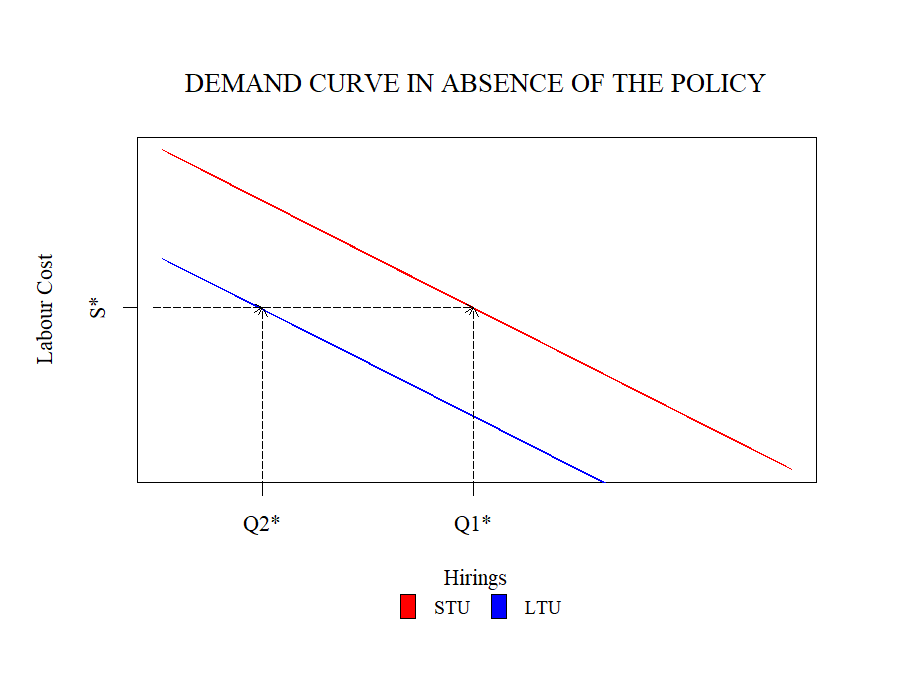}\\
Note that the shape of the two curves is simplified for illustrative porpoises. As it is possible to see, in correspondence of the same labour cost, the demand of LTU is always lower than the demand of short term unemployed. The difference between the two hirings amount is due to the lower desirability of LTU. Hence it can be interpreted as the combination of the effects of the lack of recent working experience, of the reduction of contacts with the labor market and of the scarring effect of LTU. Note that, even though LTU are less desirable than other unemployed, their curve of demand isn't constantly equal zero. This is coherent with empirical evidence. Indeed, LTU are still hired. Moreover, it is coherent with literature on scarring effect, from which it emerged the last was heterogeneous across individuals, type of job or contracts (\cite{Omo1997}, \cite{Bae2014}, \cite{Ayl2013}). The fact that the curve of demand of LTU isn't constantly equal zero suggests employers group to be highly heterogeneous. Some employers, depending on their characteristics, their experiences and on the type of work they offer, may not stigmatize LTU. \par

\subsection{Law 407/90 and Generalised Incentives}\label{targnontarg}
Law 407 was promulgated on December 29th 1990. According to it, any firm had access to tax credits for a period of 36 months, at the condition of hiring, with a permanent contract, individuals who had been either in unemployment status, or suspended from their job, or in Cassa Integrazione (temporary layoff), for at least 24 months. In Italy, firms have to pay, for each employee, a rate of her wage to the social security service, and a much smaller rate to an institution providing work insurance. The tax credits corresponded, for regular firms, to 50\% of the total amount of taxes the firm would have had to pay for the hired individuals. It corresponded to 100\% of the same amount for artisans firms and firms located in the Mezzogiorno area of Italy. The policy aimed at reducing the rate of long-term unemployed and workers in Cassa Integrazione or suspended.  In order to avoid the temptation, for the employers, to substitute workers of the firm with individuals hired through law 407/90, a firm, to be eligible, couldn't have experienced, in the last six months, firings or workers' suspending or voluntary resignation, or the end of a temporary contract. In June 28th 2012 the rules defining eligibility on firm side were relaxed and the conditions to be classified as unemployed slightly changed. The law ended on December 31st 2014 according to 2015 Legge Stabilità, promulgated on December 23rd 2014. \par
We focused on the impact of this policy on LTU hirings, not considering the other categories in the targeted group. This policy was widely exploited on italian labour market. According to INPS' reports it has been the policy, for the increase of permanent contracts, with the highest number of recipients between 2011 and 2014. In particular, the number of individuals who benefitted by this Law went from a minimum of 295'417 to a maximum of 305'327 among the years 2011-2014 (where this amount includes suspended workers and workers in temporary layoff as well)\footnote{Source: Statistiche in breve, Politiche Occupazionali e del Lavoro, \href{url}{http://servizi.inps.it/banchedatistatistiche/menu/index.html}}. In the following graph it is represented the number of individuals with respect to their unemployment duration from 2011 to 2014. \par
\includegraphics[scale=0.4]{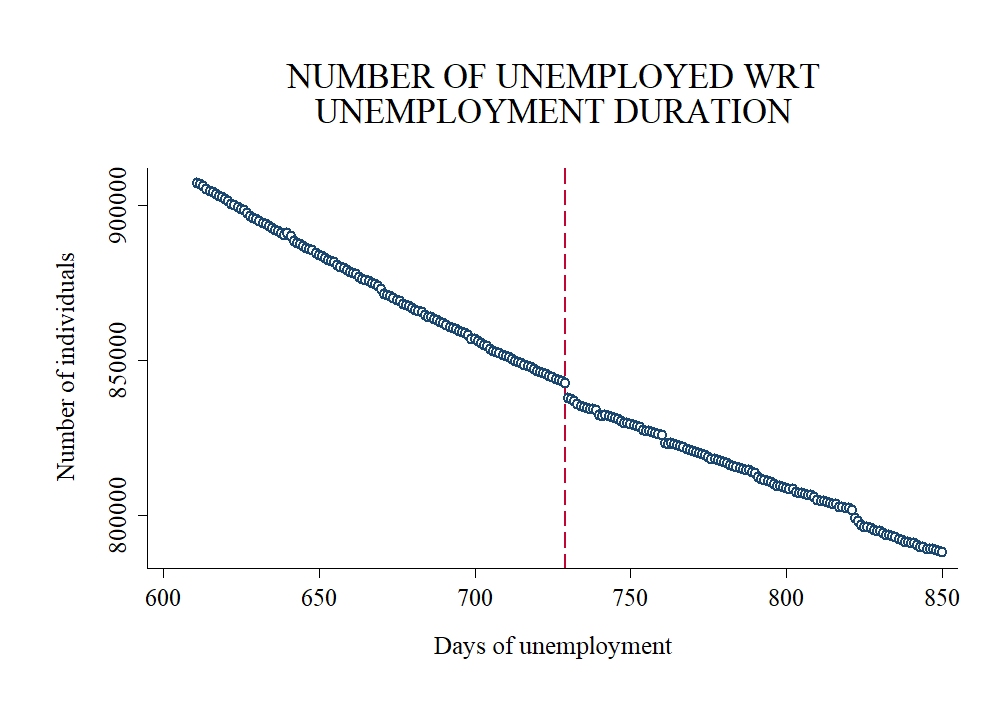}\\
It is possible to see a clear downward jump at the 729 threshold defining eligibility (evidenced by the red line). This gives a first suggestion on policy effectiveness.\\

The generalised incentives provided by Legge Stabilità 2015, consisted in tax credits to firms hiring unemployed with a permanent contract or turning a temporary contract to a permanent ones. Tax credits corresponded to 100\% of the taxes the firm had to pay to the social security service for each individual. In order to avoid the temptation, for the employers, to fire and followingly hire again their employees, firms couldn't obtain the tax credit hiring individuals who were fired from a permanent contract less than six months before. The policy, implemented since January 1st 2015, lasted one year\footnote{It was followed by an equivalent policy with tax credits corresponding to 50\% of labour taxes.}. The development of this policy raised a huge debate on the effectiveness of the law. Indeed, its implementation was used as a political weapon from all the political fronts. The policy was widely used. According to INPS' data, 1'078'885 were the new permanent contracts stipulated taking advantage from it\footnote{Source: Osservatorio Precariato, INPS}. In the following graph it is represented the share of hired individuals among unemployed with slightly less than 24 months of unemployment across months for 2014 and 2015. \par
\includegraphics[scale=0.4]{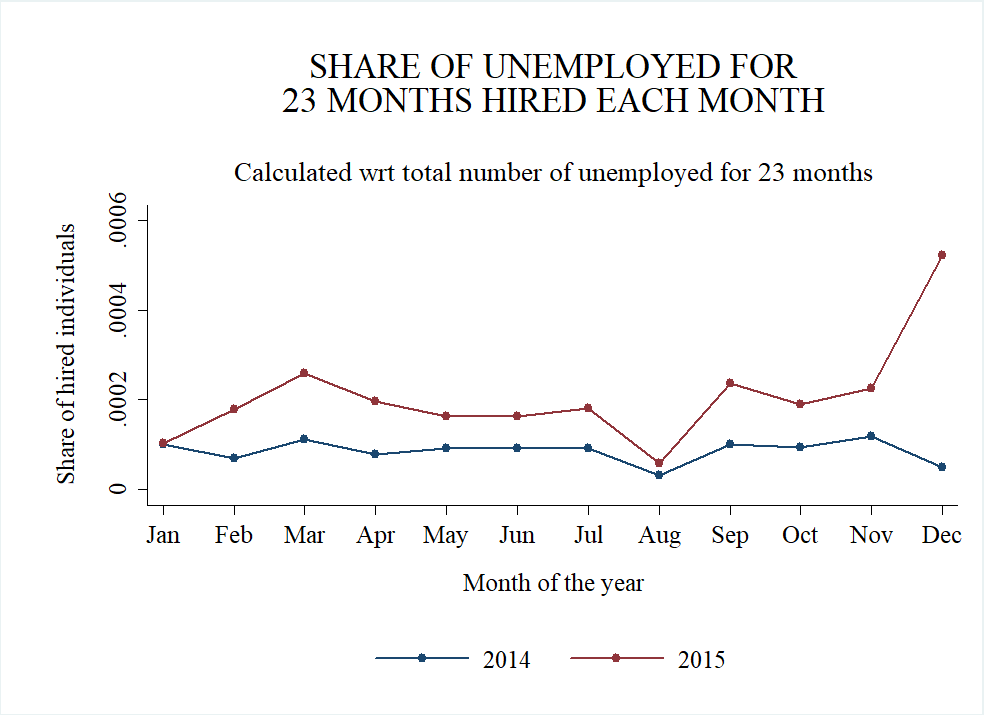}\\
It is possible to see that hirings in 2015, hence under Law 190 incentives, are higher than in 2014, in absence of incentives (indeed the group was not eligible for Law 407/90 subsidies). This suggests Law 190 incentives were effective. It is important to notice the peak in hirings in December 2015. This peak can be attributed to the fact that the policy was implemented for a limited and short period of time and its duration was communicated in advance. To compare the two types of incentives in a more general way, we estimated the impact of the untargeted incentives excluding the observation about December 2015. \\

The two policies don't differ only with respect to their target group. Indeed, while Law 407/90 gave 100\% tax credit only to determined firms, 2015 Legge Stabilità gave it to all firms. Moreover, Legge Stabilità 2015 incentives covered  the taxes due to the social security service only, while Law 407/90 covered both the taxes due to the social service and the taxes due to the institution providing work insurance (the last are significantly smaller\footnote{The percentage of wage payed to the social security service is, on average, 29.8\%. The percentage of wage payed to the insitution providing work insurance is, on average 2.9\%}). These differences appear negligible with respect to the target difference. Indeed, the restriction of the analysis to the Mezzogiorno area (where both Law 407/90 and Law 190 gave 100\% tax credits) doesn't give significantly different results (see appendix \ref{AppE}). Moreover, a comparison between the estimated average amount of subsidies given under Law 407/90 in the period 2011-2014 and the average amount that would have been given under Law 190 for the same individuals and the same wages suggests they don't differ much economically (see appendix \ref{AppA}). Another source of difference between the two laws could be the consciousness of firms about the existence of the policies. Nevertheless, the wide use of Law 407/90 and the attention given by the media to 2015 Legge Stabilità suggests both the policies were well known by italian firms. \par

\subsection{Two Types of Hiring Subsidies}
The main difference between targeted and untargeted incentives is that the first lower the labour cost in relative terms, while the second lower them only in absolute terms. Hence, if targeted incentives are implemented the LTU become economically more convenient than STU. This is not the case with untargeted incentives. The idea behind targeted incentives is that the lower labour cost of the targeted group is enough to counteract its lower desirability. Empirical evidence suggests in the heterogeneous employers group there will be some for whom the incentive isn't enough to counteract the lower desirability. Indeed, under the policy, LTU hirings are still lower than STU's. This group of employers, won't react differently to the two different types of incentives. Equally, the group of employers who doesn't consider LTU as less desirable won't react differently. In both cases, the first category of employers won't hire LTU and the second category will hire LTU considering only their lower labour cost. The eventual presence of a third group of employers will make the difference between the two incentives types. There may be a group of employers for whom the LTU have a lower desirability but for whom the relatively lower labour cost under Law 407/90 is enough to counteract it. If this group exists, its hirings will be higher under Law 407/90 than under 2015 Legge Stabilità incentives. Indeed, under the first policy the lower desirability of LTU is counteracted by the relatively lower labour cost. Under the second policy, instead, the labour cost of STU is lowered as well, hence in relative terms the labour cost of LTU doesn't change.\\
As mentioned before, the main difference between the two incentives we studied is that Law 407/90 ones targeted LTU, while 2015 Legge Stabilità incentives were extended to all unemployed. This element, and the proximity of the two policies in time, allow us to take general conclusions on the difference between targeted and untargeted incentives from the comparison between their impacts. 

\section{Data and Sample Definition}
Thanks to an agreement between Dipartimento di Scienze Sociali ed Economiche of Sapienza University of Rome and INAPP (Istituto Nazionale per l'Analisi delle Politiche Pubbliche) we had access to the used data. We used a micro-databases, available, through INAPP, to the participants to the agreement. It is an administrative database, called CICO. It was provided by the Italian Ministry of Labour and Social Policies (Ministero del Lavoro e delle Politiche Sociali). It contains all recorded employment and parasubordinate contracts\footnote{The last is a type of contract, present on italian labour market, having some of the characteristics of employment and some of the characteristics of self-employment} and some self-employment events (coming from INPS' data) for a random sample of individuals. Each record corresponds to a different contract and reports the worker ID, the firm ID, contract's and job's characteristics and starting and ending dates, and some basic socio-demographic characteristics of the individual, such as age, income and region of residence. The data are collected by the Ministry directly from the employers, who must register the contract and provide all the informations. After the collection of records, the last are submitted, by the Ministry, to a validation procedures. The data started to be recorded from 2008. Nevertheless, previous observations were reconstructed by the Ministry to provide additional informations. These data have some great advantages. From 2008 on, they are precise and valid from the point of view of records of contracts' start and end. They report detailed working histories of the individuals, in the continuous across nine years. Nevertheless, they have some limits. Given that the data report only employment experience, there is a lack of informations on individual's status in the period between the end of a contract and the start of the subsequent ones. Hence, we can't determine with certainty whether the individual was actually unemployed in the missing spell \footnote{I.e. other categories not covered by CICO informations besides unemployed are students, housekeepers, inactive retired, inactive disables, etc.}. Nevertheless, we took into account of this issue in the analysis (see section \ref{sipuo}).\par

Using CICO, we built a new database where each unit identifies, rather than a recorded contract, a group of individuals having a determined number of days in a non-CICO-recorded status, at a determined day. To be more precise, in the new database, unit $ij$ identifies the group of individuals with $i$ days in a non-CICO-recorded status at day $j$. With a non-CICO-recorded status we mean an employment status not recorded in CICO database (for simplicity we'll call it ``non-occupation'' from now on, since most of the occupational statuses are recorded). This means unit $ij$ contains informations about the group having the last recorded contract ending $i$ days before day $j$ and the next starting after, or on, day $j$. Consequently, each unit starts to be counted from the end of the first contract recorded on CICO. This re-elaboration of the data has a great advantage. Indeed, it allows to exploit the continuus nature of the data, maintaining a number of observations and variables low enough for the analysis to be computationally feasible. Without the data aggregation, in order to do the analysis across all days, we would have had to register 1825 variables reporting the employment status of 3'138'373 individuals. With the aggregation the number of observations to register and process is 6'357'570 but only one variable is needed. At the same time, the aggregation doesn't cause any loss of useful informations.\par
Given that the records are not reliable if recorded before 2008, we only considered records subsequent to December 31st 2007. Since we mostly focused on long-term unemployed this means we have useful informations starting from 2010. Indeed, it is from 2010 that we start having units far enough from the last recorded contract to be possibly considered LTU. The variables recorded in the new database for each unit $ij$ are the share of individuals, in the group, that are hired in day $j$, the total number of individuals in the group and the share of individuals in the group with given socio-demographic characteristics.\\

Following \textcite{Sch2013}, \textcite{Ana2013}, \textcite{Boo2012}, \textcite{Hut2013}, \textcite{Ham2008}, \textcite{For2004}, \textcite{Ses2016} and \textcite{Cen2016}, we used eligibility to define the treated and control groups. This means we estimated intention-to-treatment effect, rather than average treatment effect. Following \textcite{Sch2013}, we can report three main reasons to do it. The first one is that comparing workers with similair probability to be treated, don't assure the absence of selection bias in the estimation of the average treatment effect. Indeed, the subsidy has to be required by the employer and there may be an employer selection. Hence, controlling for employees characteristics only, may not be enough to solve entirely the selection bias problem. The second is that, exploiting the eligibility threshold of the policy we can use a RDD. Finally, using eligibles as treated group, the estimated effect is the intention-to-treatment ones, the last is the main parameter of interest from a policy perspective. Indeed, wage subsidies can't be mandatory. Wage subsidies' policies give the possibility to take up a subsidy, not the subsidy itself. Policy makers only have control on the intention to treatment, they have no power on subsidy reception. It is of biggest interest the impact of policies policy makers can control, and the last have control only on eligibility rules. A fourth reason, strictly linked with our framework, is that, using subsidy reception as treatment, it would have been hard, to determine how the implementation of 2015 Legge Stabilità affected the treated group. \\

In order to estimate the ITT of Law 407/90 we used data about the period 2011-2014 concerning the control and treated group chosen according to bandwidth selection method. To be able to determine the presence of indirect effect we considered all units with a forcing variable lower than the threshold across the period 2011-2016. Finally, in the regression discontinuity design for the incentives of Law 190 impact, we used records for the control group across the period 2010-2015. \par  

\section{Identification Strategy}
\subsection{Law 407/90 Intention-to-Treatment Effect}\label{mod}
To estimate the impact of Law 407/90 we applied a regression discontinuity design with daily fixed effects. It has been demonstrated that, when units at the threshold are considered, the regression discontinuity design is, at the threshold, as reliable as the golden standard of policy evaluation: randomized treatment assignment (\cite{Lee2008}). \par
The main condition to apply this method is the presence of a threshold, defined on a continuous variable, and defining treatment assignment. In this context, eligibility (hence our treatment) is defined with respect to the 24 months threshold on the continuous variable reporting unemployment days. \par
We used the following linear regression model:
\begin{equation}
\begin{array}{c}
y_{ij}=\alpha+\beta D_{ij}+\theta_{j}+\epsilon_{ij}
\end{array}
\end{equation}
Where the dependent variable is the share (with respect to the number of individuals in the entire group) of individuals of group $(ij)$ hired in day $j$. Variable $D_{ij}$ is a dummy taking value 1 if individuals in group $(ij)$ are eligibles. The $\theta_{j}$ are the daily fixed effects. The presence of daily fixed effects allow us to consider each daily comparison as independent from the others. Moreover, it allows us to overcome seasonality issues. Contracts' starting and ending dates are characterised by a strong seasonality. Not taking into account of it could bias the estimation of treatment effect. To be more clear, there are months of the year, as December, characterized by a high number of contracts' endings and a low number of hirings.  Since individuals become eligible in the month their last contract ended, there is a big group of individuals becoming eligible in a month characterized by low hirings and this can bias the estimation. Controlling for the hiring day we overcome this issue. It is easy to see that the effect of Law 407/90 is given by parameter $\beta$. \par

We followed \textcite{Cat2015} and \textcite{Li2015} approach to select the bandwidth. Its basic idea is to include in the estimation only units close enough to the threshold, to be able to consider the model as a local randomized experiment. We used  \textcite{Cat2015} methodology modifying it. To exploit the fact that we have a particularly high number of observations around the threshold when doing inference and to take into account of daily fixed effects included in the model when implementing the balancing test. Given the high number of observations around the threshold, we didn't need to apply the methodology proposed by \textcite{Cat2015} to do inference, avoiding the functional form assumptions it requires. To do their balancing test, we used a paired t-test rather than a regular ones, pairing units belonging to the same day. This guaranteed us the treated and control groups to be balanced and comparable each single day. In the balancing test, we used as covariates the share, in the group defined from the unit, of women, of individuals with different educational qualification, of individuals who started their first job at different age class, of foreign citizens, of individuals working in different working sectors or geographical areas during the last contract. The approach we applied to select the bandwidth allowed us not to include the forcing variable in the model. The correctness of its exclusion was confirmed by the fact that, inside the bandwidth, the forcing variable has no significant impact on the outcome for treated and control group separately. The selected bandwidth went from 714 to 744 days of unemployment (where 729 is our threshold). Hence, we included all units becoming eligibles in two weeks or less and all units who became eligibles since two weeks or less.\par
The results of this estimation are reported in table \ref{tab2} (for obvious reason we excluded the daily parameters). 
\begin{table}[h!]
\centering
\begin{tabular}{lc} \hline
VARIABLES & Coefficients \\ \hline
 &  \\
Treat & 3.01e-05*** \\
 & (4.62e-06) \\
Constant & 9.38e-05*** \\
 & (2.82e-06) \\
 &  \\
Observations & 45,291 \\
 R-squared & 0.094 \\ \hline
\multicolumn{2}{c}{ Robust standard errors in parentheses} \\
\multicolumn{2}{c}{ *** p$<$0.01, ** p$<$0.05, * p$<$0.1} \\
\end{tabular}
\caption{Intention to treatment effect of Law 407/90 is given by Treat.}
\label{tab2}
\end{table}
As common, we use, as significativity level, a 0.05 ones. The results suggest Law 407/90 had a positive and significant impact. Even though the magnitude of the impact may look small this is not the case. Indeed, the outcome itself has a very low magnitude. Comparing the estimated impact with the average weighted outcome of the control group, we can conclude the policy increased the last by 36\%. This value is bigger than the results obtained in previous literature. Most of the studies estimating intention-to-treatment effect concluded there wasn't a significant effect (see \cite{Boo2012}, \cite{Hut2013}, \cite{Sch2013}) and one of them found a positive and significant effect of 10\% in the short run (\cite{Ham2008}). Nevertheless, this difference can be explained by three reasons. The first one is that the policy we studied was widely used. The second is that it was implemented across a very long period (i.e. 24 years). The third is that almost all of the afore-mentioned authors (with the exception of \cite{Sch2013}) used a methodology which estimated the impact of the policy for the whole population of eligibles. Using a RDD, instead, we obtained only local results. This means our estimation describes the effect of the policy only for the group of eligibles included in the bandwidth. This is one of the most affected groups. Indeed, of the hirings under Law 407/90, around 38\% involved individuals with 24 to 27 months of unemployment. \\

\subsubsection{Bandwidth Choice in a time-varying forcing variable framework}
The approach to use in bandwidth selection has been led by the nature of our forcing variable. The last has the particularity to change over time. Consequently we can't widen the bandwidth and use the linear polynomial approach. To explain the differences between a time fixed and a time varying forcing variable we have to take into account of that, and its consequences. In standard contexts where regression discontinuity design is implemented, there is a single event where a single forcing variable (and, consequently, the treated group) is defined. In this context, instead, a unit having a determined value of the forcing variable at day $j$ will have a different forcing variable at day $j+1$. In standard contexts, the forcing variable of a unit can be considered as a resume of all the characteristics determining its selection into treatment. In the case of a time-varying forcing variable, instead, all units, regardless of their characteristics, will have a forcing variable of 0 days at the start of their unemployment period, most of them will have a forcing variable of 5 days, and so on so forth. We can then imagine there are two types of forcing variable: the potential and the observed ones. The potential forcing variable resumes the characteristics of the unit determining its selection into treatment (in our framework the potential length of its unemployment spell at hiring time in absence of treatment). The observed forcing variable is the forcing variable we're able to observe. In standard framework, the observed forcing variable and the potential forcing variable coincide. Hence, it is enough to control for the observed forcing variable to be sure to control for all the characteristics determining selection into treatment. In time-varying forcing variable frameworks, instead, the observed forcing variable takes multiple values and it doesn't necessary coincide (i.e. it coincides one day only in our framework) with the potential forcing variable. Nevertheless, there is some sort of correlation among the two forcing variables. Indeed, it is unlikely for an individual with a potential forcing variable of 3 months to have an observed one of 24 months. In particular, it is reasonable to hypothesize a group of units with an observed forcing variable of $FV_{Obs}=f$ to have a distribution of the potential forcing variable with few units presenting a value of $FV_{Pot}<<f$ and most of the units whose potential forcing variable is equally distributed in the interval $FV_{Pot} \in [f-\epsilon, \infty)$ with $\epsilon\geq 0$ reasonably small. As an example, the distribution of the potential forcing variable across groups with an observed forcing variable of 3, 23 or 24 months can be hypothesized to be as represented in the following graph\footnote{Notice that the hypothesis we're making on potential forcing variable distribution is more stringent than what we need. Indeed, for our considerations to be valid it is enough that groups of units with close values of the observed forcing variable have similar distribution of the potential ones, while groups of units with remote values of the observed forcing variable don't.}. Note that this representation is fully hypothetical. 
\includegraphics[scale=0.6]{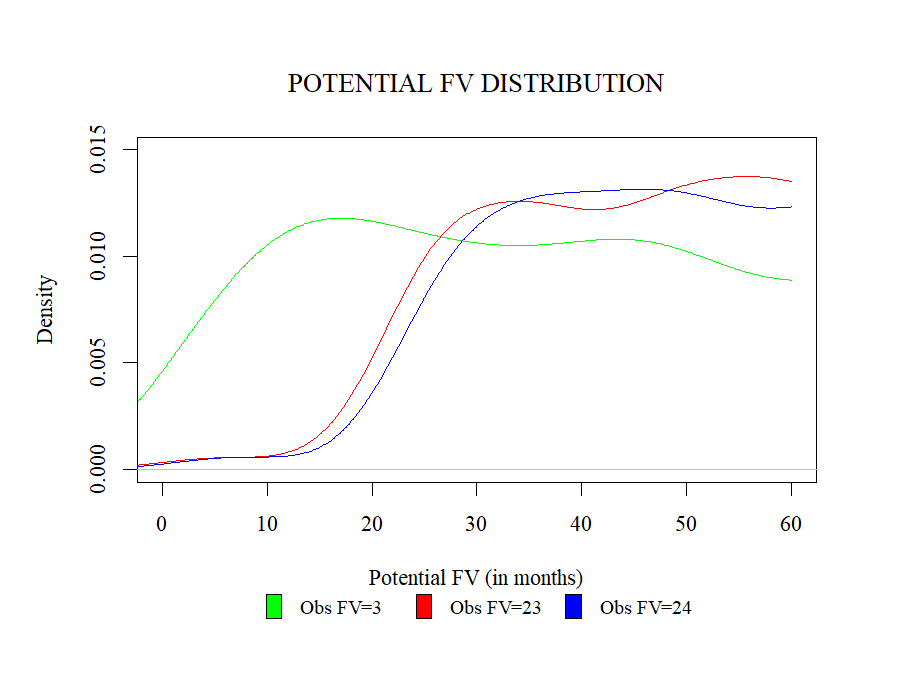}\\
It is possible to see that under this reasonable hypothesis, groups of individuals with similar values of the observed forcing variable have similar distributions of the potential ones. Hence, if the bandwidth on the observed forcing variable is chosen to be small enough, it is reasonable to assume that treated and control groups have the same potential forcing variable distribution. And, consequently, the same characteristics in terms of selection into treatment determinants. \textcite{Cat2015} method allows to choose a bandwidth with these characteristics. Indeed, it allows to choose a bandwidth such that, the group of units having an observed forcing variable under the threshold doesn't differ from the group having it over the threshold in terms of selection-into-treatment-determinant characteristics. On the contrary, the approach based on local polynomial, which use a much larger bandwidth, wouldn't guarantee us that the treated and controls have the same characteristics. \\

\subsubsection{Are the Assumptions to use Regression Discontinuity Design Plausible in this Context?} \label{sipuo}
Three main assumptions are required in order to use the Regression Discontinuity Design. The first one is the assumption of randomness in the distribution of units around the threshold. It is likely to be satisfied. Indeed, being, at a determined day, unemployed from 23 months and 28 days or 24 months, is not under control of the unemployed. There is a random component on the day the previous contract ended. There is another random component in the employers the unemployed got in contact with and the time it happened (to be unemployed since 24 months when hired, the unemployed has to find the right match at the right moment). The second one is the requirement of absence of sorting due to forcing variable manipulation. This is certainly satisfied. Indeed, the definition of the forcing variable determining law eligibility was based on the registration of the unemployed, to the employment agency, at the unemployment period starting date. Hence, the value of the individual forcing variable was completely under control of the employment agency. The unemployed had no control on it. The third assumption, namely the stable unit treatment value assumption, may be violated. Nevertheless, we checked it taking into account of displacement effect (see \ref{displ}) and controlling for the reception of other benefits in sample selection. \\
The use of CICO database requires an additional assumption. In CICO, there are no informations precisely on unemployment. This can cause a biased estimation if the control and the treated groups behave differently during the non-CICO-detected periods. As an example, if most controls are studying while most treated are actually unemployed during the 24 months, this can cause higher hirings of the first group. The consequence would be an underestimation of the policy impact. Nevertheless, this problem only arises if the distributions of non-CICO-detected spells is different between the treated and the control groups used in the estimation. This is unlikely to be, given the bandwidth selection criteria. Nevertheless, we checked for it looking to the distribution of individuals among different non-CICO-detected categoris. We used RTFL data. The last are survey quarterly data from ISTAT. Using them, we were able to verify that the distribution of individuals between different categories of non-CICO-detected spells was similar between treated and control groups each year (see appendix \ref{AppC}). \\
The model was robust to changes in the bandwidth, to covariates addition and to placebo tests both with respect to different thresholds and with respect to different years (see appendix \ref{AppD}). To have an additional proof of the validity of our methodology, we studied the correlation between the number of subsidised individuals and the estimated treatment effect across different years. In table \ref{tab7}, the coefficients of the treatment and the number of subsidised are reported. \\
\begin{table}[h!]
\centering
\begin{tabular}{ccc}
\hline
\multicolumn{2}{c}{Correlation} & 0.98808\\
\hline
Year & Treatment Effect & Subsidised \\
\hline
& & \\
2011	&	4.34e-05	 &	25202	\\
2012	&	4.69e-05	&	26539	\\
2013	&	2.48e-05	&	20118	\\
2014	&	3.81e-05	&	24689	\\
\hline
\end{tabular}
\caption{Correlation between the number of subsidised individuals and ITT across years.}
\label{tab7}
\end{table}
The number of subsidised and estimated treatment effect present a high correlation. This suggests our regression model effectively detected the impact of the studied policy. \\

\subsection{Indirect Effects}\label{displ}
Taking into account of the indirect effects of a policy is crucial for two reasons. First of all, it gives a deeper overview on policy effects. Secondly, it allows to check the correctness of the estimated impact. Indeed, if the indirect effects affect the control group the estimation is biased. In their studies, \textcite{Bro2012} and \textcite{Cal2002}, underlined as targeted hiring subsidies may have different indirect side effects.  Among the effects enumerated, two are relevant in our setting. The first one is that non-targeted individuals having similar characteristics to the targeted ones, may be penalised by policy implementation. Indeed, thanks to the policy, they have similar characteristics, but a higher labour cost, than targeted individuals. In our setting this means that hirings of individuals with, g.e. 23 months of unemployment may decrease when the policy is implemented because firms prefer to hire eligibles when hiring someone who's unemployed from long time. This side effect is called displacement effect. The second side effect is due to the presence of asymmetric information. Indeed, the public authorities providing the incentives have not perfect information on all firms and unemployed. This means firms and unemployed can cheat in order to get the subsidy even when they're not eligibles. In our framework, i.e., firms can decide to hire an individual when she is close and under the threshold defining eligibility (as a g.e. when she is unemployed from 23 months) and wait to hire her until she becomes eligible. We can call this effect post-poned hirings effect. \\
Both displacement and post-poned hirings effects imply a reduction in hirings of uneligible individuals whose value of the forcing variable is close (and obviously under) the threshold. They don't affect units with a value of the forcing variable too far from the threshold. I.e., it is unlikely displacement effect affects individuals with, as an example, 13 months of unemployment. Indeed, the two groups have completely different characteristics. Similarly, it is unlikely that an employer would accept to wait 11 months, until the 13 months unemployed becomes eligible, to hire her, just to get the subsidy. This is the only observable element we can use to determine whether the two effects are present (and, eventually, correct the counterfactual estimation for them). To use it, we can't just compare the outcome for units close to and under the threshold with the outcome for units far\footnote{When we say units ``far'' we mean they are far enough to reasonably assume they are not affected by displacement and post-poned hiring effects.} from and under the threshold. Indeed, their outcome would probably be significantly different in absence of the policy as well. Hence, we compared the difference between the outcome after 2014 and the outcome before 2015 for units close to and under the threshold with the difference in the correspondent outcomes for units far from and under the threshold. We exploited the fact that Law 407/90 ended in 2014, hence there can't be displacement and post-poned hiring effects in 2015 (at least not for the group we checked for). If these two effects are present, we would expect the size of the difference to be smaller far from the threshold than close to the threshold. In absence of them, instead, there would be no reason for the difference to diverge getting closer to the threshold. For this method to be valid, we have to assume that, in absence of the policy, the difference would be smooth across close intervals of the forcing variable. This allows us to attribute the absence of regularity in the difference to the displacement and post-poned hirings effects.\\
In the following image, is it possible to see the Kernel-weighted local polynomial smoothing of the difference between the average outcome before 2015 and after 2014 with respect to the forcing variable. This first graphical analysis suggests there are no post-poned hirings and displacement effects. \\
\includegraphics[scale=0.45]{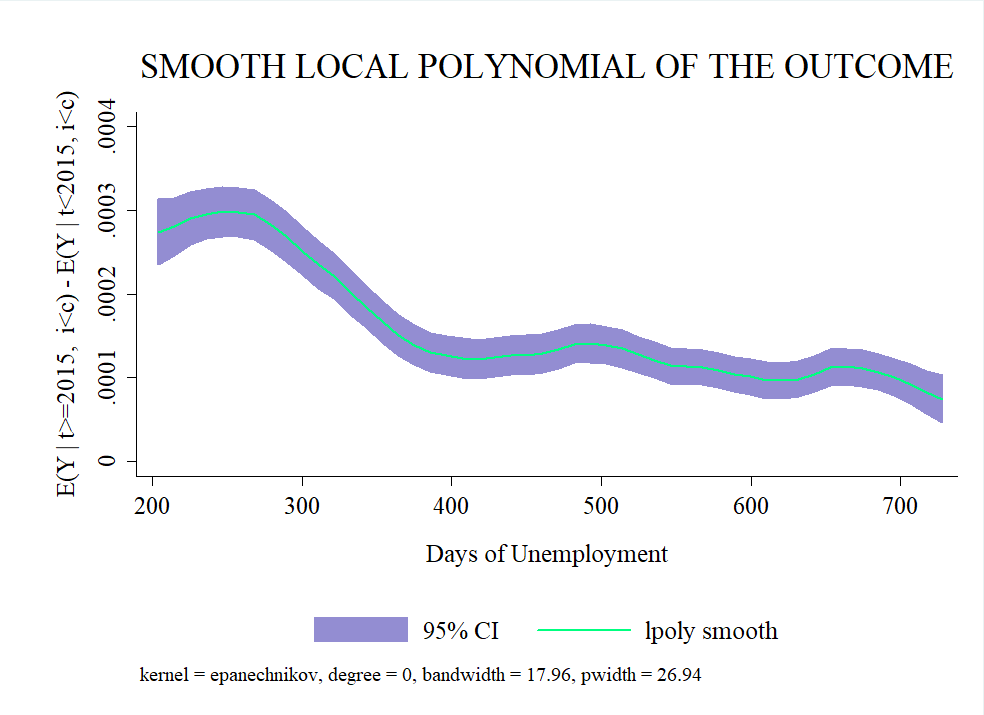}\\
Indeed, even though the difference decrease close to the threshold (where the threshold corresponds to a forcing variable value of 729), it does it far from the threshold as well. There is no unusual behaviour getting closer to it. \par
This conclusion is consistent with the results of the statisical tests comparing the average difference between two different groups of units far from the threshold and one group of units close to it. In tables \ref{tab3} and \ref{tab4} the results of some Welch t-tests are presented. The difference for the interval defining the control group is compared with the difference for two intervals defined on values of the forcing variable far from the threshold.\\
\begin{table}[h!]
\centering
\begin{tabular}{ccccc}
\hline
 & \multicolumn{4}{c}{Comparison group: [365:380] Days of Unempl}\\
\hline
Year & Diff Mean & Std. Err. & CI 95\% Lower & CI 95\% Upper\\
\hline
2011	&	-2.19e-06	&	5.91e-05	&	-1.18e-04	&	1.14e-04	\\
2012	&	1.288e-04	&	7.23e-05	&	-1.32e-05	&	2.707e-04	\\
2013	&	 -1.17e-05	&	4-69e-05	&	-1.039e-04	&	8.04e-05	\\
2014	&	-9.62e-07	&	4.61e-05	&	-9.15e-05	&	8.96e-05	\\
\hline
\end{tabular}
\caption{Welch t-tests results: Checking for the presence of Displacement and Post-Poned Hirings Effects.}
\label{tab3}
\end{table}
\\
\begin{table}[h!]
\centering
\begin{tabular}{ccccc}
\hline
 & \multicolumn{4}{c}{Comparison group: [545:560] Days of Unempl}\\
\hline
Year & Diff Mean & Std. Err. & CI 95\% Lower & CI 95\% Upper\\
\hline
2011	&	-3.64e-06	&	6.17e-05	&	-1.247e-04	&	1.175e-04	\\
2012	&	1.138e-04	&	5.8e-05	&	-4.96e-09	&	2.276e-04	\\
2013	&	8.91e-05	&	6.04e-05	&	-2.94e-05	&	2.076e-04	\\
2014	&	1.033e-04	&	5.52e-05	&	-4.96e-06	&	2.116e-04	\\
\hline
\end{tabular}
\caption{Welch t-tests results: Checking for the presence of Displacement and Post-Poned Hirings Effects.}
\label{tab4}
\end{table}
From the tests' results it is possible to see that displacement and post-poned hirings effects are not detected in any year. \\
The fact that these effects aren't present may look surprising. Nevertheless, two elements have to be considered. The first one is that often employers prefer to hire individuals as soon as possible. As an example, from a study by \textcite{Obe2008}, it emerged that they often prefer to hire short term unemployed rather than employed individuals, because the first are immediately available to work. Equally,  they may prefer to hire uneligibles close to the threshold immediately, rather than wait for them to become eligibles. In other words, they may prefer a worker immediately available rather than the subsidy. The second one, is that in Italy there are mainly small and medium firms. A small firm won't probably receive a huge amount of applications and curricula at a time. Hence, it is unlikely the firm receives the curriculum of an individual close and under the threshold and the curriculum of an individual close and over the threshold at the same time. Just as it is unlikely, when it receives a curriculum from an individual close and under the threshold, that the firm will decide not to hire her and wait for the curriculum of an eligible individual. Hence, the absence of displacement effect isn't surprising as well. This last result is in contrast with \textcite{Ann2008} study and \textcite{Cal2002} literature review. Indeed, according to the first there was a displacement effect damaging short term unemployed. Nevertheless, her analysis was based on a theoretical model which strongly simplified reality. According to the second, the displacement effect of targeted employment subsidies had an average effect of 84\%. Nevertheless, the value was based on a single study consisting in surveys to employers. If an employer answer ``yes'' when someone asked her whether she would prefer to hire a subsidised individual with respect to an unsubsidised individual with similar characteristics, this doesn't mean if she actually receive the curriculum of the unsubsidised she will reject it and wait for another curriculum by an eligible with similar characteristics. The result on the absence of post-poned hirings effect, instead, is consistent with previous literature and, in particular, with \textcite{Boo2012} and \textcite{Sch2013} conclusions.\\

\subsection{Generalised Incentives Intention-To-Treatment Effect}\label{modelloII}
To estimate the impact of generalised hiring incentives from a LTU perspective, we used another regression discontinuity design.  We used time as a forcing variable and exploited the January 1st 2015 threshold defining eligibility. We couldn't estimate the impact of the policy on the exact same group targeted by Law 407/90. Indeed, we wouldn't be able to distinguish between the effect of Law 407/90 ending and the effect of the generalised incentives start. Hence, we estimated the intention-to-treatment effect on a control group selected according to the modified \textcite{Cat2015} method (see \ref{mod}) and assumed it not to differ significantly from our target group. Given that we used monthly dummies we needed the control group to be balanced with the LTU group at a monthly level. Hence, we used the paired t-test pairing units belonging to the same month and year (nevertheless, the results don't change in a significant way if we do the same analysis on the control group used in \ref{mod}). The assumption that the two groups don't differ significantly is necessary to compare the intention-to-treatment effects of the two policies. In order to estimated the impact of untargeted incentives generally rather than the impact of Law 190 incentives themselves we excluded observations about December 2015. Indeed, the peak corresponding to that observations is probably due to a characteristic specificly of this policy: its short duration. We want more general results, representing the impact of a generic untargeted policy and being comparable with Law 407/90. We applied the following regression model to data on the period from January 2010 to November 2015:\\
\begin{equation}
y_{k}=\alpha+\sum_{l=1}^{12}\theta_{l}m_{l,k}+\gamma_1T_k+\gamma_2T_k^2+\gamma_3P_k+\gamma_4T_k*P_k+\gamma_5T_k^2*P_k+\epsilon_k
\end{equation}
We aggregated the units, with values of the forcing variable inside the selected bandwidth, at a daily level. In the model, $m_{l,k}$ are dummies taking value 1 if unit $k$ belongs to month $l$ and zero otherwise. The monthly dummies are used to take into account of seasonality. $T_k$ is the time variable and $P_k$ is a dummy flagging the period Law 190 was in place. The rest of the notation is as before. The effect of the policy is given by the coefficient $\gamma_3$. We decided not to use a higher order polynomial to avoid overfitting (\cite{Gel2017}). Table \ref{tabsenzadic} reports the result of this estimation (for brevity porpoises we excluded the monthly dummies).\\

\begin{table}[h!]
\centering
\begin{tabular}{lc} \hline
VARIABLES & Coefficients \\ \hline
 &  \\
Time & -6.26e-08*** \\
 & (2.27e-08) \\
Time$^2$ & 2.47e-11** \\
 & (1.10e-11) \\
Policy & -6.74e-03 \\
 & (4.96e-03) \\
Policy*Time & 6.79e-06 \\
 & (4.99e-06) \\
Policy*Time$^2$ & -1.69e-09 \\
 & (1.25e-09) \\
Constant & 1.59e-04*** \\
 & (1.57e-05) \\
 &  \\
Observations & 2,160 \\
 R-squared & 0.087 \\ \hline
\multicolumn{2}{c}{ Robust standard errors in parentheses} \\
\multicolumn{2}{c}{ *** p$<$0.01, ** p$<$0.05, * p$<$0.1} \\
\end{tabular}
\caption{Generalised incentives ITT on vulnerable group is given by $\gamma_3$. Observations about December 2015 excluded.}
\label{tabsenzadic}
\end{table}
It is possible to see that the intention-to-treatment effect of the incentives is non-significant. The estimated impact is not different from the ones estimated in the placebo tests (see section \ref{placebo}). We can conclude the effect of these incentives is significantly smaller than Law 407/90 ones. As said before (see section \ref{targnontarg}), this difference is mainly attributable to the fact that Law 407/90 incentives are targeted and Law 190 incentives are not. This result suggests the impact of targeted incentives is entirely due to the lower labour cost of the targeted group, which counteract its lower desirability. Consequently, when a generalised policy is implemented, it should be accompained by additional incentives tailored around the vulnerable groups of unemployed in order for them to be affected by it. \par

To provide additional informations on the mechanism behind the policies and on Law 190 incentives impact, we repeated the analysis including observations about December 2015. The results of this estimation are reported in table \ref{tab8}. 
\begin{table}[h!]
\centering
\begin{tabular}{lc} \hline
VARIABLES & Coefficients \\ \hline
 &  \\
Time & -6.26e-08*** \\
 & (2.28e-08) \\
Time$^2$ & 2.49e-11** \\
 & (1.11e-11) \\
Policy & 0.0211** \\
 & (8.99e-03) \\
Policy*Time & -2.15e-05** \\
 & (9.09e-06) \\
Policy*Time$^2$ & 5.48e-09** \\
 & (2.29e-09) \\
Constant & 1.46e-04*** \\
 & (1.63e-05) \\
 &  \\
Observations & 2,191 \\
 R-squared & 0.126 \\ \hline
\multicolumn{2}{c}{ Robust standard errors in parentheses} \\
\multicolumn{2}{c}{ *** p$<$0.01, ** p$<$0.05, * p$<$0.1} \\
\end{tabular}
\caption{Law 190 incentives ITT on vulnerable group is given by $\gamma_3$.}
\label{tab8}
\end{table}
\\
The policy had a positive and significant impact. The last was considerably lower than Law 407/90 impact. This suggests a significant component of the impact of Law 190 incentives is probably due to the limited duration of the policy itself. In particular, the whole effect of the incentives can be attributed to the December peak in hirings. 

\subsubsection{Are the Assumptions to use RDD with time forcing variable Plausible in this Context?}
The use of models based on time discontinuity requires some conditions to be satisfied. First of all there can't be an anticipation effect. If policy effect is anticipated with respect to policy implementation starting time, the last can't be used as threshold in the discontinuity analysis. Luckily, the policy was announced on December 23rd 2014, few weeks before its implementation starting time. It is therefore unlikely there was an anticipation effect. \textcite{Sou2002} suggested other assumptions which have to be satisfied when models based on time discontinuity are used. In particular, there shouldn't be shocks happening at the policy time and having a significant impact on the outcome, other than the policy itself. Possible causes of violation of this condition are an exogenous changement in overall economic conditions, the implementation, in the same period, of other laws, or a possible changement in the composition of the study population. Considering the first source of violation, in our framework, an increase in hirings may follow, as an example, an improvement in overall economic conditions. To verify whether this is the case, we added to the regression three exogenous variables that can proxy the economic conditions of the Country and observed that the intention-to-treatment effect estimation didn't change (suggesting the assumption holds). The first two were the annual final consumptions of non-resident families in Italy both in current and one year lagged values. The third one was the one year lagged quarterly value of GDP. The use of the lagged value is required for the variable to be exogenous and it is justified by the fact that, as mentioned by \textcite{Cen2016}, the improvement in the economic conditions usually have a delayed effect on hirings. We used ISTAT data on GDP and consumptions\footnote{Data Sources: \href{url}{http://dati.istat.it/}}. \\
The second possible source of violation comes from the fact that in 2015 the Jobs Act was implemented. This Law brought significant changes to permanent contracts' rules, relaxing the costraints a permanent contract implied on employer's side starting from March 2015. To verify whether this brought to an overestimation of Law 190 intention-to-treatment effect, we re-estimated the model excluding observations from March 2015 on. The new estimated intention-to-treatment effect was higher than the ones obtained excluding December 2015 only (and it was still non significant). This suggests Jobs Act impact was too small to affect the estimation of Law 190 incentives ITT. This result is in line with \textcite{Ses2016} study. From it, it emerged that the effect of Jobs Act's changements on permanent contracts rules was negligible with respect to the effect of Law 190.\par
A third possible source of violation could be a change in the composition of population study. To verify whether that was the case, we included in the regression model covariates on educational qualification and employment sector shares and observed that the ITT estimation didn't change significantly. Another possible source of bias was a changement in the share of unemployed in non-occupied group happening at policy implementation time. To take into account of it, we included in the regression model, a variable measuring the yearly rate of unemployed among non-CICO-detected individuals and observed that this didn't change ITT estimation. The variable was built using RTFL data. The results of all the analysis described in this section are presented in appendix \ref{AppF}.\\

\section{Conclusions}
The previous analysis allows to make several considerations. Nevertheless, making them, it is important to remember that the results of this analysis are only local, given the method we used. The following conclusions should be attributed only to the groups included in the analysis. \par
Law 407/90 had a significant and strong intention to treatment effect on eligibles with approximately 24 months of unemployment. Indeed, its implementation increase their likelihood to be hired by 36\%. Additionally, the policy didn't present side effects as displacement effect and post-poned hirings effect. This is probably due to the fact that employers prefer to hire the chosen workers immediately and to italian firms characteristics. \par
A generalisation of the incentives to all unemployed would strongly penalise the vulnerable group of LTU. Indeed, the intention-to-treatment effect of Law 190 parcel out of the component due to the limited implementation period of the policy is non significant. This, and \textcite{Cen2016} and \textcite{Ses2016} results, suggest that the generalisation of the policy re-allocate the benefits in favour of the other, more ``desirable'', groups of unemployed. In order to be effective, with respect to vulnerable groups of unemployed, a policy based on hiring subsidies should lower their relative labour costs rather than their absolute ones. This would avoid the benefits redistribution.\par
Thanks to the peak in hirings of LTU in December 2015, probably due to the limited duration of the policy, Law 190 incentives had a positive and significant effect on LTU hirings. Nontheless, the effect was considerably lower than Law 407/90 ones.

\nocite{*}
\addcontentsline{toc}{section}{Bibliography}
\pagestyle{plain}
\printbibliography

\appendix
\section{\\Estimation for the Mezzogiorno Area}\label{AppE}
In the following are presented the results of the analysis reduced to the group of individuals having the last working experience in a region of the Mezzogiorno Area. In table \ref{tab21} the results of the estimations of treatment effect of Law 407/90 is reported together with its value under different robustness checks. As mentioned before, the estimation is fairly similar to the one obtained for Italy as a whole. Moreover, it is robust to all checks. The selected bandwidth, in this analysis, was 718 to 740 days of unemployment. \par
\begin{table}[h!]
\centering
\begin{tabular}{lcccccc} \hline
 & (1) & (2) & (3) & (4) & (5) & (6) \\
THRES & Model 1 & Model 2 & Model 3 & Model 4 & Model 5 & Model 6 \\ \hline
 &  &  &  &  &  &  \\
24 & 4.72e-05*** & 4.30e-05*** & 2.73e-05*** & 1.99e-05 &  &  \\
 & (9.01e-06) & (9.12e-06) & (1.02e-05) & (1.55e-05) &  &  \\
22 &  &  &  &  & -4.64e-06 &  \\
 &  &  &  &  & (8.10e-06) &  \\
26 &  &  &  &  &  & -9.00e-06 \\
 &  &  &  &  &  & (8.91e-06) \\
 &  &  &  &  &  &  \\
Obs & 33,603 & 33,580 & 27,390 & 12,782 & 43,378 & 40,595 \\
 R-sq & 0.068 & 0.071 & 0.093 & 0.192 & 0.059 & 0.078 \\ \hline
\multicolumn{7}{c}{ Robust standard errors in parentheses} \\
\multicolumn{7}{c}{ *** p$<$0.01, ** p$<$0.05, * p$<$0.1} \\
\end{tabular}
\caption{Law 407/90 ITT estimation and robustness check for the Mezzogiorno Area. Model 1: Standard. Model 2: Covariates addition. Model 3: Different bandwidth [722:736]. Model 4: Different bandwidth [726:732]. Model 5 and 6: Placebo Tests .}
\label{tab21}
\end{table}

\begin{table}[H]
\centering
\begin{tabular}{lcccccc} \hline
 & (1) & (2) & (3) & (4) & (5) & (6) \\
THR & Model 1 & Model 2 & Model 3 & Model 4 & Model 5 & Model 6 \\ \hline
 &  &  &  &  &  &  \\
01/01/2015 & -5.03e-03 & -5.89e-03 & -4.85e-03 & -6.04e-03 & -3.19e-03 &  \\
 & (5.56e-03) & (5.56e-03) & (5.54e-03) & (5.61e-03) & (5.62e-03) &  \\
01/01/2014 &  &  &  &  &  & -8.18e-04 \\
 &  &  &  &  &  & (2.08e-03) \\
 &  &  &  &  &  &  \\
Observations & 2,160 & 2,160 & 2,160 & 1,795 & 2,160 & 1,826 \\
 R-squared & 0.085 & 0.086 & 0.090 & 0.096 & 0.102 & 0.040 \\ \hline
\multicolumn{7}{c}{ Robust standard errors in parentheses} \\
\multicolumn{7}{c}{ *** p$<$0.01, ** p$<$0.05, * p$<$0.1} \\
\end{tabular}
\caption{Generalised incentives ITT on vulnerable group estimation and checks for the Mezzogiorno Area. Model 1: Standard. Model 2: Lagged GDP addition. Model 3: Foreign Consumption addition. Model 4: Percentage of unemployed among non-CICO detected addition. Model 5: Educational Level and Sector variables addition. Model 6: Placebo test.}
\label{tab22}
\end{table}
Assumptions and robustness checks suggest the model is valid. The intention-to-treatment effect of Law 407/90 is slightly higher in the Mezzogiorno area than in the rest of Italy. Nevertheless, the Mezzogiorno area estimation presents a high volatility, hence the two coefficients are not significantly different. There is almost no difference in terms of generalised incentives intention-to-treatment effect. \par

\section{\\Comparison of the estimated averages of subsidies amounts under the two policies}\label{AppA}
Given that the rules for the determination of benefit amount are well known and that we had informations about employees' wages, we were able to compare the amount of subsidies given under the two laws. To do it, we used informations on wages to estimate approximatively\footnote{Where it is approximatively because the rate of wage the taxes have to be payed on, depends on the type of work and the sector. Hence, we used an average value.} the tax credit amount given to each individual, hired under Law 407/90, under the two policies. We compared the average value of the estimated amounts each year. In theory, Law 407/90 should imply higher benefits, since it covers both social security service and the amount due to the institution providing work insurance, while Law 190 covers only the first. Nevertheless, the rate of wage due to the institution providing work insurance is very low. Hence, the difference may be negligible. Moreover, Law 407/90 covers 100\% of the taxes only for the Mezzogiorno area or for artisans firms, balancing even more the average amount. The comparison is reported in table \ref{tab50}.
\begin{table}[h!]
\centering
\begin{tabular}{cccc}
\hline
Year & Avg 407 & Avg 190 & Diff /Avg 190\\
\hline
2010 & 7023 & 5726 & 0.227 \\
2011	&	5816	&	5479	&	0.061	\\
2012	&	5846	&	5502	&	0.063	\\
2013	&	5722	&	5378	&	0.064	\\
2014	&	6821	&	6366	&	0.071	\\
\hline
\end{tabular}
\caption{Comparison between estimated average subsidies under the two policies.}
\label{tab50}
\end{table}
As expected, the difference between the average amount of subsidy given under each policy is negligible in most years. Indeed, the difference is lower than 7.2\% of the smallest average almost all years, with the exception of 2010. For this reason observations relative to that year have been excluded from the estimation. \par

\section{\\Comparison of the distribution of treated and controls among non-CICO-detected categories}\label{AppC}
In the following graph are represented the distributions, among different non-CICO-detected categories, of individuals with 23 and individuals with 24 months of unemployment. In order to build these distributions we used RTFL data. The last is the result of a quarterly survey conducted by ISTAT (Italian National Institute of Statistics) to provide informations on labour force conditions. It is one of the most important database providing these informations and it is used to build official estimations of labour force status. Data are collected every week interviewing more than 250'000 families living in 1'100 different municipalities (\cite{Ana2016}). \par
\includegraphics[scale=0.6]{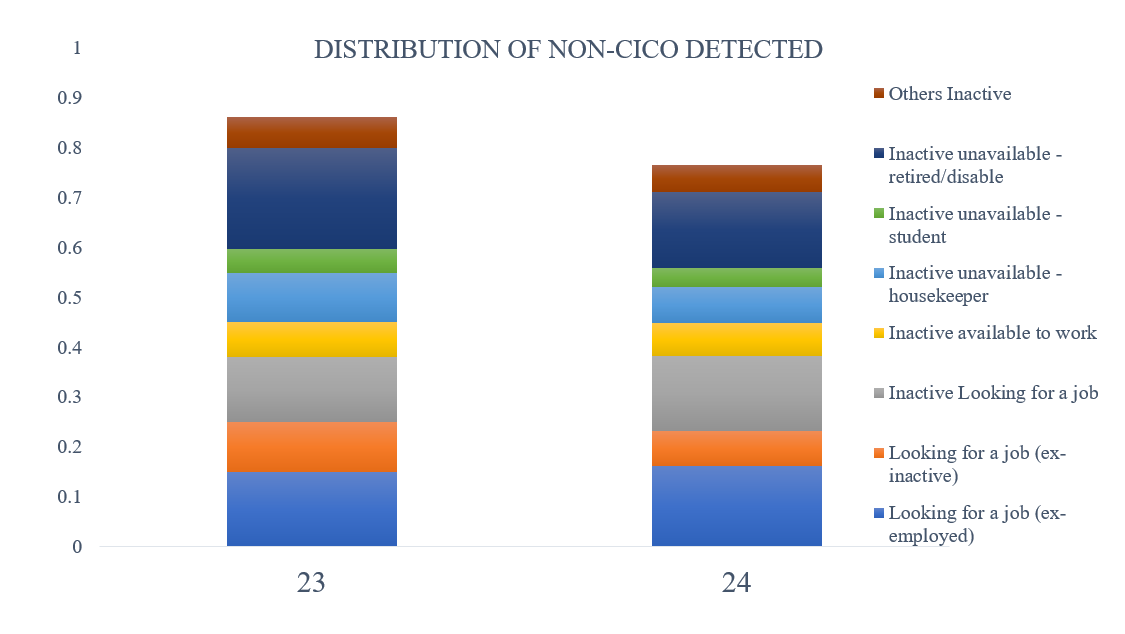}\\
The two distributions are fairly similair. Given the reduced amount of observations with the required unemployment period we weren't able to obtain the informations for the exact groups used in the estimation. Nevertheless, given the methodology we used to select the bandwidth, the difference in the distributions should be even smaller in the selected groups.

\section{\\Placebo and Robustness Checks, Law 407/90 ITT Estimation}\label{AppD}
\subsection{Placebo Test}
\begin{table}[H]
\centering
\begin{tabular}{lccc} \hline
 & (1) & (2) & (3) \\
THRESHOLD & 2011-2014 & 2011-2014 & 2015 \\ \hline
 &  &  &  \\
22 Months & 9.87e-08 &  &  \\
 & (4.86e-06) &  &  \\
26 Months &  & -7.88e-06 &  \\
 &  & (5.04e-06) &  \\
24 Months &  &  & -9.60e-06 \\
 &  &  & (1.28e-05) \\
 &  &  &  \\
Observations & 42,369 & 42,369 & 11,315 \\
 R-squared & 0.077 & 0.109 & 0.189 \\ \hline
\multicolumn{4}{c}{ Robust standard errors in parentheses} \\
\multicolumn{4}{c}{ *** p$<$0.01, ** p$<$0.05, * p$<$0.1} \\
\end{tabular}
\caption{Placebo Tests.}
\label{tab51}
\end{table}

\subsection{Robustness Check with respect to Covariates Addition}
\begin{table}[H]
\centering
\begin{tabular}{ccc}
\hline
VARIABLES & Coefficients & Std Error \\
\hline
$\alpha$  & -1.43e-05 & (1.416e-04) \\
Treatment ($\beta_1$) & 3e-05*** & (4.59e-06) \\
Women \% & 8.64e-05 & (8.27e-05) \\
Education \% (Elementary baseline) &  & \\
Low Secondary  & -5.48e-05 & (9.44e-05) \\
Up Secondary & 1.329e-04 & (1.18e-03) \\
Tertiary no Univ & -1.099e-03* & (6.436e-04) \\
Tertiary Univ & -4.33e-05 & (1.215e-04) \\
Degree+ & -1.252e-03** & (5.514e-04) \\
1st Job Age (15-19 baseline) & & \\
20-24 & -3.26e-05 & (1.416e-04) \\
25-29 & 6.86e-05  & (1.279e-04) \\
30-44  & 1.189e-04  & (1.214e-04) \\
45+ & -9.32e-05 & (1.701e-04) \\
Foreigners & -7.47e-05 & (1.228e-04) \\
Sectors \% (Agricolture baseline) & & \\
Industry & 1.66e-04 & (1.097e-04) \\
Constructions & 3.128e-04***  & (7.71e-05) \\
Services & 4.8e-05  & (6.52e-05) \\
Job Area \% (NW baseline) & \\
NE & -1.104e-04  & (1.085e-04) \\
Center & 1.6e-05  & (1.148e-04) \\
South and Islands & 7.26e-05 & (8.55e-05) \\
 &  & \\
Observations & 45,291 & \\
 R-squared & 0.096 & \\ \hline
\multicolumn{3}{c}{ Robust standard errors in parentheses} \\
\multicolumn{3}{c}{ *** p$<$0.01, ** p$<$0.05, * p$<$0.1} \\
\end{tabular}
\caption{Robustness check: Covariates addition.}
\label{tab52}
\end{table}

\subsection{Bandwidth Changes}
\begin{table}[H]
\centering
\begin{tabular}{lccc} \hline
 & (1) & (2) & (3) \\
VARIABLES & [714:744] & [720:740] & [724:734] \\ \hline
 &  &  &  \\
Treat & 3.01e-05*** & 2.43e-05*** & 2.21e-05*** \\
 & (4.62e-06) & (5.75e-06) & (8.22e-06) \\
 &  &  &  \\
Observations & 45,291 & 30,681 & 16,071 \\
 R-squared & 0.094 & 0.124 & 0.204 \\ \hline
\multicolumn{4}{c}{ Robust standard errors in parentheses} \\
\multicolumn{4}{c}{ *** p$<$0.01, ** p$<$0.05, * p$<$0.1} \\
\end{tabular}
\caption{Bandwidth Changements.}
\label{tab51}
\end{table}
Note that the different bandwidths we tested for, are all smaller than the chosen bandwidth. Indeed, given the method we used in bandwidth selection, in a bandwidth bigger than the chosen ones there wouldn't be balance between treated and controls. The estimation of treatment effect is not very sensitive to bandwidth selection. 

\section{\\Placebo and Robustness Checks, Generalized Policy ITT Estimation}\label{AppF}
\subsection{Placebo Test}\label{placebo}

\begin{table}[H]
\centering
\begin{tabular}{lcccccc} \hline
 & (1) & (2) & (3) & (4) & (5) & (6) \\
VAR & Model 1 & Model 2 & Model 3 & Model 4 & Model 5 & Model 6 \\ \hline
 &  &  &  &  &  &  \\
Treat & -6.74e-03 & -6.37e-04 & -7.29e-04 & 3.50e-05 & 3.70e-05 & 5.24e-05 \\
 & (4.96e-03) & (1.62e-03) & (1.59e-03) & (2.42e-04) & (2.40e-04) & (7.02e-05) \\
 &  &   &  &  &  &  \\
Obs & 2,160 & 1,826 & 1,826 & 1,826 & 1,826 & 1,826 \\
 R-sq & 0.081 & 0.057 & 0.057 & 0.057 & 0.057 & 0.057 \\ \hline
\multicolumn{7}{c}{ Robust standard errors in parentheses} \\
\multicolumn{7}{c}{ *** p$<$0.01, ** p$<$0.05, * p$<$0.1} \\
\end{tabular}
\caption{Placebo Tests. Model 1: Standard. Model 2: January 1st 2014 as threshold. Model 3: December 31st 2013 as threshold. Model 4: January 1st 2013 as threshold. Model 5: December 31st 2012 as threshold. Model 6: January 1st 2012 as threshold.}
\label{tab52}
\end{table}

\subsection{Checking for other Exogenous Shocks}
\begin{table}[H]
\centering
\begin{tabular}{lccccccc} \hline
 & (1) & (2) & (3) & (4) & (5) & (6) & (7) \\
VAR & Model 1 & Model 2 & Model 3 & Model 4 & Model 5 & Model 6 & Model 7 \\ \hline
 &  &  &  &  &  &  &  \\
Treat & -6.74e-03 & -7.16e-03 & -6.54e-03 & -6.77e-03 & -6.90e-03 & -1.95e-03 & -1.34e-04 \\
 & (4.96e-03) & (4.96e-03) & (4.93e-03) & (4.95e-03) & (4.98e-03) & (2.25e-03) & (2.78e-04) \\
 &  &  &  &  &  &  &  \\
Obs & 2,160 & 2,160 & 2,160 & 2,160 & 1,795 & 1,885 & 2,160 \\
R-sq & 0.087 & 0.087 & 0.094 & 0.089 & 0.097 & 0.059 & 0.111 \\ \hline
\multicolumn{8}{c}{ Robust standard errors in parentheses} \\
\multicolumn{8}{c}{ *** p$<$0.01, ** p$<$0.05, * p$<$0.1} \\
\end{tabular}
\caption{Checking for other exogenous shocks. Model 1: Standard. Model 2: Lagged GDP addition. Model 3: Foreign Consumption addition. Model 4: Lagged Foreign Consumption addition. Model 5: Percentage of unemployed among non-CICO detected addition. Model 6: Exclusion of observations relative to Jobs Act implementation months. Model 7: Educational Level and Sector variable addition.}
\label{tab53}
\end{table}

\end{document}